\documentclass[12pt]{article}

\usepackage{graphicx}
\usepackage{amsmath}
\usepackage{amsthm}
\usepackage{amsfonts}
\usepackage[utf8]{inputenc}
\usepackage{latexsym}

\newtheorem{teor}{Theorem}[section]
\newtheorem{prop}{Proposition}[section]

\newtheorem{cor}{Corollary}[section]
\newtheorem{obs}{Remark}[section]
\newtheorem{defin}{Definition}[section]
\newtheorem{exem}{Example}[section]

\newtheorem{algor}{Algorithm}[section]

\newfont{\Mb}{msbm10}
\newcommand{\C}{\mbox{\Mb\symbol{67}}}
\newcommand{\R}{\mbox{\Mb\symbol{82}}}

\begin{document}
\setcounter{equation}{0}
\setcounter{figure}{0}
\setcounter{table}{0}

\hspace\parindent
\thispagestyle{empty}

\bigskip        
\bigskip 
\bigskip

\begin{center}
{\LARGE \bf A Linear Prelle-Singer method}
\end{center}

\bigskip

\begin{center}
{\large
$^a$L.G.S. Duarte, $^a$H.S. Ferreira $^a$L.A.C.P. da Mota,\footnote{E-mails: lgsduarte@gmail.com, higorferreira@unifei.edu.br and lacpdamota@gmail.com}
}

\end{center}

\bigskip
\centerline{\it $^a$ Universidade do Estado do Rio de Janeiro,}
\centerline{\it Instituto de F\'{\i}sica, Depto. de F\'{\i}sica Te\'orica,}
\centerline{\it 20559-900 Rio de Janeiro -- RJ, Brazil}

\bigskip
\bigskip
\bigskip
\bigskip

\abstract{The Prelle-Singer method allows determining an elementary first integral admitted by a polynomial vector field in the plane. It is a semi-algorithm whose nonlinear step consists of determining the Darboux polynomials of the vector field. In this article we construct a linear procedure to determine the Darboux polynomials present in the integrating factor of a polynomial vector field in the plane. Next, we extend the procedure to deal with rational 2ODEs that admit an elementary first integral. 
}
 
\bigskip



\bigskip

\section{Introduction}
\label{intro}

Determining first integrals of vector fields is an old and quite complicated problem. With the advent of computer algebra systems (CAS), the area enjoyed a `renaissance' in the second half of the 20th century\footnote{Despite this, even when dealing with Liouvillian first integrals of polynomial vector fields in the plane, it was only in 2019 that G. Chèze and T. Combot developed a semi-algorithm for the general case \cite{ChCo}.} \footnote{See \cite{Sin,Chr,Nosjpa2001,Nosjpa2002-2,ChGiGiLl,Nosjcam2005,ChLlPaWa1,FeGi,Che,BoChClWe,Zha,ChLlPaWa5,ChCo,Nosjde2021,DeGiVa,Nosjmp2009} and references therein.} and, in particular, we can highlight the work of M. Singer who, in 1983 (together with M. Prelle), presented a work \cite{PrSi} whose results allowed the implementation of a semi-algorithm that, given a plane polynomial vector field $\mathfrak{X}$, could decide whether or not it admits an elementary first integral and, if so, determine it. The fundamental result establishes that, {\it If a polynomial vector field in the plane admits an elementary first integral, then it presents an algebraic integrating factor formed by rational powers of eigen-polynomials (Darboux polynomials) of the vector field}. In practical terms, the main task in the Prelle-Singer (PS) method consists of determining Darboux polynomials (DPs), that is, polynomials $p$ such that $\mathfrak{X} (p) = q\,p$, where $q$ is a polynomial called cofactor\footnote{The method is a {\it semi}-algorithm because we do not have a bound for the degree of DPs present in the integrating factor and, thus, the procedure may never finish.}. The fact that the `eigenvalue' $q$ is a polynomial instead of a constant makes the determination of the DPs a nonlinear process and, because of this, the complexity grows very quickly with the increase in the degree of DPs present in the integrating factor\footnote{If we use the method of undetermined coefficients (MUC), for DP degrees greater than 4 the problem may be computationally unfeasible in practice.}.

In this work, in section 2, we built a semi-algorithm ({\it LPS}) to determine, with a linear process, the DPs $p_i$ present in the integrating factor of a polynomial vector field in $\R^2$ (our main result). With the exception of singular cases, the procedure is general and very easily implementable. In section 3, we extend the {\it LPS} algorithm in order to determine (linearly) the DPs present in an integrating factor of rational second-order ordinary differential equations (rational 2ODEs) that admit an elementary first integral. Finally, we present a brief discussion and some comments on the efficiency and applicability of the method.

\section{A linear PS method}
\label{lpsm}

\subsection{The `original' Prelle-Singer method}
\label{opsm}

Consider the polynomial dynamical system 
\begin{equation}
\label{sd2}
\left\{ \begin{array}{l}
\dot{x} = N(x,y), \\
\dot{y} = M(x,y),
\end{array} \right.
\end{equation}
where $M$ and $N$ are coprime polynomials in $\C[x,y]$. A first integral $I$ of the system (\ref{sd2}) is a function that is constant over the solutions of (\ref{sd2}).

\begin{defin}
Let $E$ be an elementary field extention of $\C(x,y)$. A function $I \in E$ is said to be an {\bf elementary first integral} of the system {\em (\ref{sd2})} if $\mathfrak{X}(I)=0$, where $\mathfrak{X} \equiv N\, \partial_x + M\, \partial_y$ is the {\bf vector field associated} (or the {\bf Darboux operator associated}) with the system {\em (\ref{sd2})}.
\end{defin}

\begin{obs}
Noting that the system {\em (\ref{sd2})} is autonomous, we can divide $\dot{y}$ by $\dot{x}$ to obtain a rational 1ODE $\,y' = {M(x,y)}/{N(x,y)}$. For this rational 1ODE, the function $I$ defines its general solution in the implicit form: $I(x,y) = c$. Besides, the 1-form $\gamma \equiv M\,dx - N\,dy$ is null over its solutions and, therefore, the exact 1-form $\omega \equiv dI$ is proportional to $\gamma$, i.e., $\omega = R\,\gamma$. We call $R$ an {\bf integrating factor}.
\end{obs}

\begin{teor}[Prelle-Singer]$\!\!:$
\label{ps}
\

\noindent
If the system {\em (\ref{sd2})} presents an elementary first integral then

\noindent
(i) It presents one of the form $I = W_0 + \sum c_k\,\ln(W_k)$, $W$'s are algebraic functions.

\noindent
(ii) The 1-form $M\,dx - N\,dy$ presents an integrating factor of the form $R=\prod_i {p_i}^{n_i}$, where the $p_i$ are irreducible polynomials and the $n_i$ are rational numbers.

\end{teor}

\noindent 
{\it Proof:} For a proof see \cite{PrSi}.

\medskip

\begin{defin}
Let $p(x,y)\,\in\, \C[x,y]$. The polynomial $p$ is said to be a {\bf Darboux polynomial} of the vector field $\mathfrak{X}$ if $\mathfrak{X}(p)=q\,p$, $q$ is a polynomial in $\C[x,y]$ called {\bf cofactor} of $p$.
\end{defin}

\begin{cor} 
\label{dppmynx}
If the hypothesis of the theorem \ref{ps} is satisfied then $\sum_i n_i \, \frac{\mathfrak{X}(p_i)}{p_i} = - \langle \nabla | \mathfrak{X} \rangle$, where $p_i$ are Darboux polynomials of $\mathfrak{X}$ and $\langle \nabla | \mathfrak{X} \rangle$ is the divergent of $\mathfrak{X}$.
\end{cor}

\begin{algor}[PS method (sketch)] \label{PSnlin} \end{algor}

\begin{enumerate}

\item Construct polynomial candidates with undetermined coefficients ($p_c$ and $q_c$) for $p$ and $q$ and substitute them in the equation $\mathfrak{X}(p_c)-q_c\,p_c=0$.

\item  Collect the equation in the variables $(x,y)$ obtaining a set of (quadratic) equations for the coefficients of the candidates. Solve this set of equations to the undetermined coefficients. 

\item Substitute the solutions in the equation $\sum_i n_i \, q_i + N_x + M_y=0$ (see corollary \ref{dppmynx}) and collect the equation in the variables $(x,y)$. Solve the set of equations for the $n_i$.

\item Construct the integrating factor $R=\prod_i {p_i}^{n_i}$ and find (by quadratures) the elementary first integral $I$.

\end{enumerate}

\medskip

\subsection{Determining DPs linearly}
\label{lpsp}

As mentioned, the nonlinear (and often problematic) step of the original PS method consists of solving the nonlinear problem $\mathfrak{X}(p)=q\,p$ (in the method outlined in subsection \ref{opsm} above, this nonlinear part corresponds to step 2). One of the keys to bypassing this non-linear step comes from the statement ({\it i}) of the \ref{ps} theorem (which is not used in the original semi-algorithm); the other key comes from the form of the partial differential equation (PDE) for the inverse integrating factor combined with its algebraic structure:

\begin{itemize}
\item From statement ({\it i}), if $\mathfrak{X}$ admits an elementary first integral, then it presents one of the form $ \,I=\mbox{\large{e}}^{W_0}\,\prod_j {W_j}^{k_j}$. In the following, for the sake of clarity and simplicity, we will treat an apparently much less general case \footnote{More about that later.}: 
\begin{prop}
\label{p1_r1sp}
Let $\mathfrak{X} \equiv N\,\partial_x+M\,\partial_y$ be a vector field, where $N,\,M \in \C[x,y]$ are coprime polynomials. Consider that $\mathfrak{X}$ admits a Darboux (non constant) first integral $\,I=\mbox{\large{e}}^{A/B}\,\prod_j {p_j}^{n_j}$ (i.e., $\mathfrak{X}(I)=0$), where the $p_i$ are irreducible polynomials, $A,\,B \in \C[x,y]$ are coprime polynomials and the $n_i$ are rational numbers. Then, in the general case, there exists an integrating factor of the form $\,R = {I}/{B^2\prod_j {p_j}}$.
\end{prop}
{\it Proof.} We have that $I_u \!=\! \left(\partial_u\left(\frac{A}{B}\right)+\frac{\sum_k n_k\,\partial_u(p_k)\,\prod_{l,l\neq k}p_l}{\prod_j p_j}\right)\,{\rm e}^{A/B}\,\prod_j {p_j}^{n_j}$, implying that

\begin{eqnarray} 
I_x\!\!\!\!&=&\!\!\!\! \frac{(A_xB-B_xA){\prod_j p_j}+B^2\sum_k n_k\,\partial_x(p_k)\,\prod_{l,l\neq k}p_l }{B^2\prod_j p_j}\,I = \frac{Pol_{[x]}\,I}{B^2\prod_j p_j}, \nonumber \\ [3mm]
I_y\!\!\!\!&=&\!\!\!\! \frac{(A_yB-B_yA){\prod_j p_j}+B^2\sum_k n_k\,\partial_y(p_k)\,\prod_{l,l\neq k}p_l }{B^2\prod_j p_j}\,I = \frac{Pol_{[y]}\,I}{B^2\prod_j p_j}. \label{ixiy}
\end{eqnarray}
So ${M}/{N} = -{I_x}/{I_y} = - {Pol_{[x]}}/{Pol_{[y]}}$ and, in the general case (i.e., $Pol_{[x]}$ and $Pol_{[y]}$ are coprime) we have that ${I}/{B^2\prod_j p_j}$ is an integrating factor.$\,\,\,\Box$

\item Second result: consider that $V$ is the inverse integrating factor of the vector field $\mathfrak{X} $ as defined in proposition \ref{p1_r1sp}. So, $V$ obeys 1PDE $N\,V_x + M\,V_y = V\langle \nabla | \mathfrak{X} \rangle$. It is a well-known fact that the solution of this non-homogeneous 1PDE can be obtained from the homogeneous 1PDE defined by\footnote{$\langle \nabla | . \rangle,\,$ stands for {\bf div}.} $N\,\psi_x+M\,\psi_y- z\,\langle \nabla | \mathfrak{X} \rangle\,\psi_z = 0$. In other words, the vector field $N\,\partial_x + M\,\partial_y - z\,\langle \nabla | \mathfrak{X} \rangle\,\partial_z$ is in the intersection of the tangent spaces $T_{1}$ and $T_{2}$ defined by the level surfaces of the functions ${\cal I}_1 \equiv I(x,y)$ and ${\cal I}_2 \equiv z\,V(x,y)$.
\end{itemize}

\begin{teor} 
\label{linPS}
Let $\mathfrak{X} \equiv N\,\partial_x+M\,\partial_y$ be a vector field, where $N,\,M \in \C[x,y]$ are coprime polynomials. Consider that $\mathfrak{X}$ admits a Darboux (non constant) first integral $\,I=\mbox{\large{e}}^{\,A/B}\,\prod_j {p_j}^{n_j}$ (i.e., $\mathfrak{X}(I)=0$), where the $p_i$ are irreducible polynomials, $A,\,B \in \C[x,y]$ are coprime polynomials and the $n_i$ are rational numbers. Consider also that the polynomials \vspace{-2mm}
\begin{eqnarray} 
Pol_{[x]}&\equiv& (A_xB-B_xA){\prod_j p_j}+B^2\sum_k n_k\,\partial_x(p_k)\,\prod_{l,l\neq k}p_l , \label{polx} \\ [1mm]
Pol_{[y]}&\equiv& (A_yB-B_yA){\prod_j p_j}+B^2\sum_k n_k\,\partial_y(p_k)\,\prod_{l,l\neq k}p_l , \label{poly} 
\end{eqnarray}
are coprime. Then the polynomial vector field $\tilde{\mathfrak{X}} \equiv \mathfrak{X}-z\,\langle \nabla | \mathfrak{X} \rangle\,\partial_z$ admits a polynomial first integral of the form $z\,\prod_j {p_j}^{k_j}$, where $p_j$ are Darboux polynomials of $\mathfrak{X}$ and $k_j$ are positive integers.
\end{teor}


\noindent
{\it Proof.} Consider that the hypotheses of the theorem are verified. From proposition \ref{p1_r1sp}, $\,V = {B^2\prod_j {p_j}}/{I}$ is an inverse integrating factor for the vector field $\mathfrak{X}$. Therefore, the functions ${\cal I}_1 \equiv I(x,y)$ and ${\cal I}_2 \equiv z\,V(x,y)$ are first integrals of the vector field $\tilde{ \mathfrak{X}}$. As any function of ${\cal I}_1$ and ${\cal I}_2$ is also a first integral, $\,{\cal I}_1\,{\cal I}_2 = z\,B^ 2\prod_j {p_j}$. $\,\,\,\Box$

\begin{cor}
\label{vpol}
If the hypotheses of Theorem \ref{linPS} are fulfilled, then $V_p \equiv B^ 2\prod_j {p_j}$ is a polynomial inverse integrating factor for $\mathfrak{X}$.
\end{cor}

\noindent
{\it Proof.} It follows directly from the proof of Theorem \ref{linPS}. $\,\,\,\Box$

\medskip
We can use the result of Theorem \ref{linPS} (and Corollary \ref{vpol}) to avoid the non-linear step of the PS algorithm, i.e., to construct a linear algorithm capable of determining the Darboux polynomials present in the integrating factor of the vector field $\mathfrak{X}$:

\begin{algor}[LPS (sketch)] \label{PSlin} 
\end{algor}
\begin{enumerate}
\item Construct the vector field $\tilde{\mathfrak{X}}$ and set a degree ($d_v$) for $V_p$. 

\item Construct a polynomial candidate $V_c(x,y)$ (with undetermined coefficients $v_c$ and degree $d_v$) for $V_p$. 

\item Substitute the polynomial candidate in the equation $\tilde{\mathfrak{X}}(z\,V_c)=0$ and collect it in the variables $(x,y,z)$ obtaining a set of linear equations for the coefficients $v_c$.

\item Solve this set of equations to the undetermined coefficients $v_c$. If there is a solution (other than the trivial one), substitute it in $V_c$ to obtain $V_p$. Otherwise increase the degree of $V_c$ ($d_v$) and go to step 2.

\item From the integrating factor $R=1/V_p$ determine (by quadratures) the first integral $I$.
\end{enumerate}

\begin{exem} {\em Consider the 1ODE given by}\label{ex1}
\end{exem} \vspace{-3mm}
\begin{equation}
\label{ex1tese1ode}
y' = {\frac {3\,{y}^{10}+18\,x{y}^{6}-9\,{x}^{2}{y}^{3}+2\ ,{x}^{3}}{{y}^{2}
  \left( -63\,{y}^{10}+51\,x{y}^{7}-7\,{x}^{2}{y}^{4}+9\,{x }^{3} \right) }}.
\end{equation}
Applying the {\it LPS} algorithm we obtain (for $d_v=13$) $V_p = \left( -3\,{y}^{3}+x \right) ^{2} \left( {y} ^{7}+{x}^{2} \right)$ in 0.016 seconds\footnote{This time interval refers to a test implementation of the algorithm in a Maple (running on a machine with intel I5 processor - 4GB - 8th Gen).}.

\section{An extension for rational 2ODEs}
\label{exten2odes}

This extension applies to 2ODEs that present two (independent) Liouvillian first integrals, implying the existence of a Darboux Jacobi multiplier (see the paper \cite{Zha} by X. Zhang). In this case, if the (Cartan) vector field $\mathfrak{X}$ associated with the 2ODE admits a polynomial inverse Jacobi multiplier, we can apply a procedure similar to that described in section 2. 

\begin{prop}
\label{ijmpol}
Let $\mathfrak{X} \equiv \partial_x+z\,\partial_y+(M/N)\,\partial_z$ be the Cartan vector field associated with the rational 2ODE $z'= \phi(x,y,z) = M/N,\, (z \equiv y')$, where $N,\,M \in \C[x,y,z]$ are coprime polynomials. If $z'= \phi(x,y,z)$ admits a polynomial inverse Jacobi multiplier $P_J$, then the vector field $\tilde{\mathfrak{X}} \equiv \mathfrak{X}-w\,\langle \nabla | \mathfrak{X} \rangle\,\partial_w$ admits a polynomial first integral of the form $w\,\prod_j {p_j}^{k_j}$, where $p_j$ are Darboux polynomials of $\mathfrak{X}$ and $k_j$ are positive integers.
\end{prop}

\noindent
{\it Proof.} It follows directly from the definition of the vector field $\tilde{ \mathfrak{X}}$. $\,\,\,\Box$

\medskip

We can extend the {\it LPS} algorithm directly using the result above:

\begin{algor}[LPS$_2$ (sketch)] \label{PSlin2} 
\end{algor}
\begin{enumerate}
\item Construct the vector field $\tilde{\mathfrak{X}} \equiv \mathfrak{X}-w\,\phi_z\,\partial_w$ and set a degree ($d_{pj}$) for $P_J$. 

\item Construct a polynomial candidate ${P_J}_c(x,y,z)$ (with undetermined coefficients $pj_c$ and degree $d_{pj}$) for $P_J$. 

\item Substitute the polynomial candidate in the equation $\tilde{\mathfrak{X}}(w\,{P_J}_c)=0$ and collect it in the variables $(x,y,z,w)$ obtaining a set of linear equations for the coefficients $pj_c$.

\item Solve this set of equations to the undetermined coefficients $pj_c$. If there is a solution (other than the trivial one), substitute it in ${P_J}_c$ to obtain ${P_J}$. Otherwise increase the degree of ${P_J}_c$ ($d_{pj}$) and go to step 2.

\item From the Darboux polynomials present in $P_J$, construct an integrating factor $R$ and (by quadratures) determine the first integral $I$.
\end{enumerate}

\begin{exem} {\em Consider the 2ODE given by}
\end{exem} \vspace{-3mm}
\begin{equation}
\label{ex2-2ode}
z' = {\frac { \!\left( {x}^{2}{y}^{3}z\!-2x{y}^{4}z\!+
{y}^{5}z\!-{x}^{2}{y}^{3}\!+2x{y}^{4}\!-\!{y}^{5}\!+3{x}^{2}yz\!-6x{y}^{2}z\!+
3{y}^{3}z\!+{y}^{2}z\!-{y}^{2}\!+\!z \right) }{ \left( x-y \right) ^{2}\left( z-1 \right) ^{-2} }}.
\end{equation}
Applying {\it LPS$_2$} we obtain (for $d_{pj}=13$) $P_J = \left( {y}^{2}z-{y}^{2}+z \right)  \left( {x}^{2}{y}^{2}z-2\,x{y}^{3}z+{y}^{4}z \right.$ $\left. -{x}^{2}{y}^{2}+2\,x{y}^{3}-{y}^{4}+{x}^{2}z-{y}^{2}z-2\,xy+2\,{y}^{2}+yz-y+2\,z-2 \right) ^{2}$ in 1.6 seconds.

\section{Discussion}
\label{discus}

Some final comments:

\begin{itemize}

\item The biggest advantages of the {\it LPS} procedure are: 1 - Linearity of the equations for indeterminates. 2 - Possibility of extension to higher orders ({\it LPS$_2$}). 3 - Simplicity of computational implementation. Regarding linearity, the efficiency of the procedures even surpasses the algorithms of G. Chèze and T. Combot in some cases: The examples cited in sections 8.2 and 8.3 of \cite{ChCo} are solved in a time interval of less than 0.016 seconds\footnote{We are referring to 2D systems (or, equivalently, rational 1ODEs) that admit an elementary first integral. It is worth remembering that Chèze-Combot algorithms are much more general.}. 

\item A closer examination of the steps 1 to 5 of {\it LPS} and {\it LPS$_2$} shows us that these procedures may never end. In other words, they are semi-algorithms.

\item The cases where $V$ (inverse integrating factor) is a non-polynomial rational function of $(x,y)$ are (contrary to our intuition) the exception and not the rule. These cases can only occur in the (very rare) cases where $Pol_{[x]}$ and $Pol_{[y]}$ (see Theorem \ref{linPS}) are not coprime, i.e., they have a common polynomial factor which is a Darboux polynomial $\overline{p}$ not present in $B^ 2\prod_j {p_j}$.

\item Among the exceptional cases mentioned in the previous comment (where $V$ is a rational function), the vast majority presents a denominator $\overline{p}$ of a much lower degree than the degree of the denominator which consists, in the most general case, of a product of Darboux polynomials ($\prod {p_l}^{m_l}$). Thus, in most cases it is much simpler to determine them. Consider, for example, the following 1ODE:
\begin{equation}
\label{ex2vnpol}
y'=-{\frac {{y}^{2} \left( {x}^{2}{y}^{4}+{y}^{3}x-1 \right) }{2\,{x}^{3}{y}^{5}+{x}^{2}{y}^{4}-2\,yx+1}}.
\end{equation}
Applying our procedure, we do not obtain a solution for any degree of $V_c$. However, applying the method of undetermined coefficients (MUC), we obtain the DP $\,y+x$ easily (since the polynomial has degree 1). Now, applying our procedure using $z\,V_c/(y+x)$ as a candidate for a first integral of $\tilde{\mathfrak{X}}$, we obtain $V= \left( {x}^{3}y+2\,{x}^{2}{y}^{2}+x{y}^{3}-1 \right) ^{2}/(y+x),$ in a practically null time interval.

\item In cases where the vector field $\mathfrak{X}$ admits a non-Darbouxian elementary first integral\footnote{Again, contrary to our intuition, this case is the exception and not the rule (see \cite{ChLlPaWa5}).}, i.e., in cases where $V^k$ ($k$ is a positive integer) is a polynomial function ($V$ is an algebraic of the form $\prod {p_j}^{n_j}, \,$ where $n_j$ are rational numbers, see \cite{PrSi}). In this case, simply use the equation $\tilde{\mathfrak{X}}(z^k\,V_c)=0$. Consider, for example, the following 1ODE:
\begin{equation}
\label{ex3vkpol}
y'=-{\frac {{y}^{2} \left( {x}^{2}{y}^{4}+{y}^{3}x-1 \right) }{2\,{x}^{3}{y}^{5}+{x}^{2}{y}^{4}-2\,yx+1}}.
\end{equation}
Applying the procedure with $k=1$, we do not obtain a solution for any degree of $V_c$. However, making $k=2$, we obtain $\,V^2 = {\left( x{y}^{2}- 1 \right) ^{3} \left( x{y}^{2}+1 \right) ^{3}}$, and therefore $V = \sqrt{\left( x{y}^{2}-1 \right) ^{3} \left( x{y}^{2}+1 \right) ^{3} }$.

\item When we have a rational 2ODE with only one Liouvillian (elementary) first integral, we can make use of the method if we are in possession of a non-local symmetry of the 2ODE (an algorithmic way of obtaining a non-local symmetry of a rational 2ODE that presents an elementary first integral is described in \cite{Noscsf2023}). In this case, simply apply the {\it LPS} algorithm to the associated 1ODE\footnote{A rational 1ODE associated with the 2ODE. See \cite{Nosjmp2009}.} to obtain the first integral.

\item When we are dealing with a 2ODE with two Liouvillian first integrals (section 3), the {\it LPS$_2$} procedure determines a Jacobi multiplier and not an integrating factor\footnote{Remember that, in the case of 2ODEs, they are different concepts.}. However, the really difficult part (determining the DPs) has already been accomplished.

\end{itemize}


\end{document}